# Detecting Exoplanets by Gravitational Microlensing using a Small Telescope

*Grant Christie*
*Auckland Observatory*
*PO Box 24-180, Auckland, New Zealand*
*grant@christie.org.nz*

**Abstract**

Gravitational microlensing is a new technique that allows low-mass exoplanets to be detected at large distances of ~7kpc. This paper briefly outlines the principles of the method and describes the observational techniques. It shows that small (e.g. 0.35m) telescopes with a CCD camera can make unexpectedly useful observations of these events. © 2005 Society for Astronomical Science.

## 1. Introduction

Auckland Observatory has been active in variable star research since 1969 when a UBV photoelectric photometer was constructed and used on the 0.5m Zeiss Cassegrain telescope. Since 2003 a CCD camera has been used and this has greatly increased the range of projects that can be undertaken.

The observatory is sited in a large park within Auckland City and suffers from the usual problems of urban light pollution. Moreover, Auckland is a relatively cloudy site suggesting that the most efficient observing strategy is to acquire data at a fast rate while the sky is clear and to give priority to projects that are tolerant of cloud interruptions.

Time series photometry of cataclysmic variable stars (CVs) provides an efficient way to use the available telescope time. Almost any observations, even those interrupted by passing cloud, can still provide a scientifically useful result. These observations are done was part of the CBA network (CBA-Auckland).

Beginning in 2003, we have also been observing gravitational microlensing events. The observational techniques are similar to those used for CVs except that the fields are always very crowded and the time scales of interest are generally longer, allowing longer exposures to be used.

In 2004 Auckland Observatory joined the MicroFUN collaboration based at Ohio State University, intrigued by the possibility of contributing to the discovery of an extra-solar planet.

Up until this time, observations of gravitational microlensing events were done by telescopes larger than 0.6m, although most were 1m or larger.

As most of these events are detected in the galactic bulge of the Milky Way, Auckland is very well placed geographically (latitude S36.9). Moreover, our time zone (GMT+12) means that we can start observing an event just as the telescopes in Chile have to stop. So while the telescope we use is small (0.35m SCT), our favorable geographic location plus the ability to schedule long hours of coverage at short notice has proved to be surprisingly successful.

This paper will introduce the basic principles of gravitational microlensing, concentrating on its application to exoplanet detection. Then we will describe the equipment, the observing protocol and the data processing techniques. Two examples of recent planet detections will be presented.

## 2. Gravitational Microlensing

The gravitational field of every star acts as a lens, deflecting the path of light passing through it. A schematic of the basic geometry is given in Figure 1.

For stellar masses, the field of view of the gravitational lens is tiny (~1mas) so the probability of seeing another star within its field of view is extremely small, typically of order $10^{-6}$.

While it is quite impractical to monitor any single lens in the hope of seeing a star pass through it, there are several well-established and efficient surveys specifically designed to detect gravitational microlensing events. This is achieved by surveying regions in the galactic bulge where star densities are very high and measuring the brightness of roughly one hundred million stars each night.

There are two microlensing surveys currently active. One is the Optical Gravitational Lensing

Experiment (OGLE) which detects ~600 events per year using 1.3m Warsaw Telescope at Las Campanas Observatory in Chile. The other is Microlensing Observations in Astrophysics (MOA) which presently detects ~50 events per year using a 0.4m telescope at Mt. John Observatory in New Zealand. A new 1.8m telescope dedicated to the MOA survey is about to start operation at Mt. John.

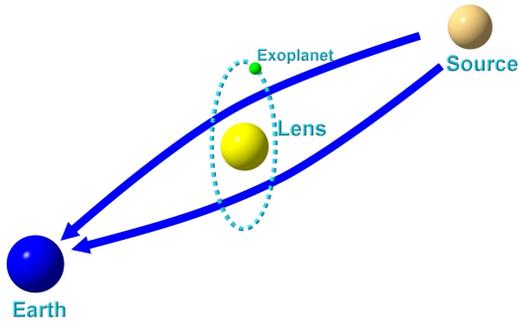

**Figure 1. The geometry of a gravitational microlensing event. The gravitational field of the lens magnifies the image of the source star and so brightens it. The presence of an exoplanet causes caustics in the lens.**

The source star is usually in the galactic bulge and hence at a distance of 6-10 kpc. The lens star may be anywhere in between the observer and the source but is typically around 2-7 kpc.

When there is perfect alignment between the source and the lens stars, the image formed by the lens of the source star is a circle, commonly called the Einstein ring. Because the Einstein ring provides a natural scale, most dimensions in microlensing are expressed in units of Einstein radius.

If the gravitational lens is formed by a single isolated star, the mathematical form of the lens magnification, A, is very simple:

$$A(u) = (u^2 + 2) / \{u(u^2 + 4)^{1/2}\} \qquad (1)$$

where u is angular separation of the source star from the center of the lens in units of the Einstein radius. The parameter, u, is commonly called the *impact parameter*. A microlensing event that conforms to this simple symmetric relationship is called a point-source point-lens (PSPL) model.

Note that the impact parameter itself is a function of time because the source star is moving through the lens. The lens magnification reaches a maximum when the impact parameter is a minimum; this occurs when the source star is closest to the center of the lens.

Depending on factors such as the relative distances of the lens and the source, it will usually take 50-100 days for the source star to cross the Einstein ring.

Any departure from this classical PSPL behavior is considered to be an anomaly and these can be caused by a variety of effects.

One fairly common anomaly is caused by the source star being a giant with a "large" angular diameter. A solar-type dwarf subtends an angular size of ~0.6µas at the galactic center so a giant star may subtend an angular size that is an appreciable fraction of the Einstein radius.

Gravitational microlensing observations may enable the diameter of the source star to be determined to very high accuracy (±0.05µas) and in favorable events, it is also possible to measure directly the limb darkening and even the oblatness of the star (Rattenbury et al, 2005).

However, the anomalies we are most interested in are those caused by planets orbiting the lensing star. When the lens is made up by two gravitating bodies, we get a more complex lens than the simple symmetric form given in Eq. (1).

In the case of a binary lens, caustic lines are formed within the lens which have some extremely useful properties. Caustics are closed contours that correspond to lines of nearly infinite magnification. If a source star passing through the lens crosses a caustic, the apparent brightness of the event increases sharply as the caustic magnifies the size of the source star.

Thus, even though the planet's mass is much smaller than the lens star, the caustics caused by the planet have a fortuitously large and directly observable effect on the event light curve - they cause a measurable anomaly (Abe et al, 2004).

To detect a planet, we only require the source star to pass over or close to a caustic. Fortunately, the planet caustics are also found close to the lens star and when the source star is close to the lens star (i.e. u is small), the magnification can become very large. At this point, the event can be bright enough to be observed with small telescopes just when there is maximum likelihood of crossing a caustic.

Thus microlensing events of very high magnification are exquisitely sensitive to the presence of planets. It also means than in order to detect a planet around the star, the event only needs to be intensively observed close to the maximum brightness – a relatively brief period of typically 1-5 days. This kind of coverage requires a network of observatories spaced in longitude and preferably at a southern latitude.

The caustics formed by a star/planet pair are a closed contour so that caustic crossings must occur as pairs – one being the entry and the other being the exit crossing. Caustic crossing are short time-scale events and because there is seldom advance warning when they are to occur, intensive photometric monitoring is required.

For massive planets similar to Jupiter, the mass ratio is $\sim 1.0 \times 10^{-3}$ whereas for Earth-mass planets, the mass ratio is $\sim 3.0 \times 10^{-6}$ (relative to the Sun). In the case of a Jovian mass planet, the probability of a detection reaches 100% over quite a large area surrounding the lens.

Once an event has been detected by a microlensing survey – that is, a *source* star is detected crossing the field of view of the *lensing* star – an alert is issued to follow-up networks which then decide if the event merits close monitoring.

The principal criterion is for the event to have a very high magnification, preferably >100. Although such cases are rare, they offer a very high probability of detecting a planet if one is present (Rattenbury et al, 2002).

Once the light curve for an event has been obtained, (albeit with inevitable gaps), a search is made for models that can accurately reproduce it. The key parameters sought are:
- the planet-star separation on the sky
- the diameter of the source star relative to that of the Einstein ring
- the angle of the source star trajectory to the planet-star axis and,
- the planet-star mass ratio.

Essentially the modeling process involves trialing very large numbers of possible trajectories of the source star through the lens, iteratively changing the free parameters while seeking an optimal match to the observational data.

For some events, the model can be determined quickly while in more complex situations it may take months of intensive effort to locate a solution.

It should be noted that many observed microlensing events follow the simple PSPL model and exhibit no anomalies. Nevertheless, even in such cases it is usually possible show there are no planets in quite broad exclusion zones around the parent star. This is valuable information that over time will help determine how many stars have planets and how many do not.

## 3. Equipment

The Nustrini Telescope at Auckland Observatory is a 0.35m Celestron Schmidt-Cassegrain manufactured in 1980. It is operated at f/11 giving a focal length of 3.85m and an image scale of 53.75 arc-sec mm$^{-1}$. The primary mirror is locked in place and focusing is effected by an electric Crayford focuser with digital readout (Jim's Mobile Inc, Lakewood CO).

The telescope mount is a Paramount GT1100s (Software Bisque, Golden CO) that is fully controlled by a computer. The absolute RMS pointing error is ~45as and periodic error is <0.5as peak-to-peak. The software control of the mount is through The Sky V5 (Software Bisque, Golden CO). Exposures of up to 600s are possible giving minimal image elongation without guiding.

The seeing in Auckland is typically in the range 2.3-3.5as with wind speeds <20km hr$^{-1}$.

The CCD camera used is an Apogee AP8p (Apogee Instruments Inc., Roseville CA) with a 1k x 1k SITe003 thinned, back-illuminated chip. It is thermoelectrically cooled and operates at –20ºC during the summer and –25ºC otherwise. This detector has high quantum efficiency and quite broad spectral response; the 24μm pixels have a well depth ~350k e$^-$. The camera is operated in 1x1 binned mode giving an image scale of 1.29as pixel$^{-1}$. The field of view is 22 arc-minutes but for time series photometry only a central subframe of 512x512 pixels is used. No filters are currently used although observing with an I-band filter may be advantageous.

The CCD camera is controlled by MaxImDL/CCD (Diffraction Limited, Ottawa, Canada).

## 4. Image Calibration

The AP8p camera bias level is nominally set at 3030ADU. However, it drifts slowly during the night so regular bias frames are taken to ensure proper dark frame subtraction. Flat fields are taken using either the dusk or dawn twilight sky whenever conditions permit. Typically, 12 full frames are taken at an altitude of 70º facing away from the sun. These are then bias subtracted, dark subtracted, normalized and median combined to produce a master flat.

Bias subtracted dark frames of long exposure time (~24hr of total dark exposure) are scaled to match the exposure time of the science frames.

The science frames are first bias subtracted and then the scaled dark frame is subtracted. Finally, the science frame is divided by the master flat.

## 5. MicroFUN

The MicroFUN collaboration is based at Ohio State University (Department of Astronomy) and has been in operation since the 2003 season. It is led by Professor Andrew Gould who is largely responsible for selecting the target list and coordinating the observing network.

The alerts issued by the OGLE or MOA surveys are monitored by MicroFUN for events favorable to planet detection. The most favorable (high magnification) events usually provide the least warning because the rise to maximum magnification might only take a few hours. On the other hand, an otherwise low interest event can suddenly become riveting if the source star happens to cross a caustic.

Events of very high magnification (>100) are rare, but they are by far the most sensitive to the presence of planets. MicroFUN tries to identify these events and then to activate an intensive observational campaign.

Each observatory in the network may be requested to halt its observing program and switch to the microlensing event at very short notice. Small observatories can usually do this without difficulty but for telescopes larger than 1m, this may not be possible because of scheduling issues. Thus, while large telescopes provide high accuracy photometry, they cannot be relied upon to be available at the critical time.

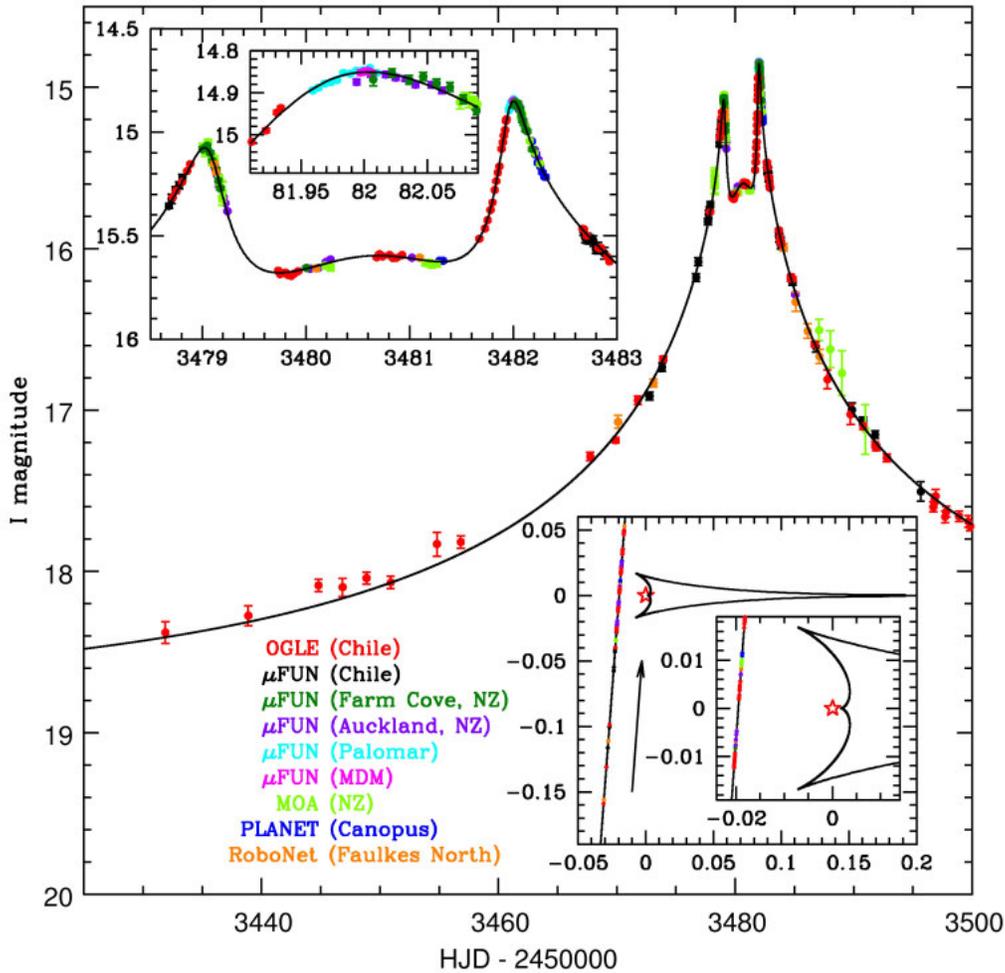

**Figure 2. The light curve of OGLE-2005-BLG-071. The upper insert shows the detail at maximum while the lower insert shows the path of the source star through the binary lens. The two peaks in the light curve are caused by the passage of the source star close to, but not actually crossing, the caustic cusps. Units are Einstein ring radius. (Udalski et al, 2005)**

*4*

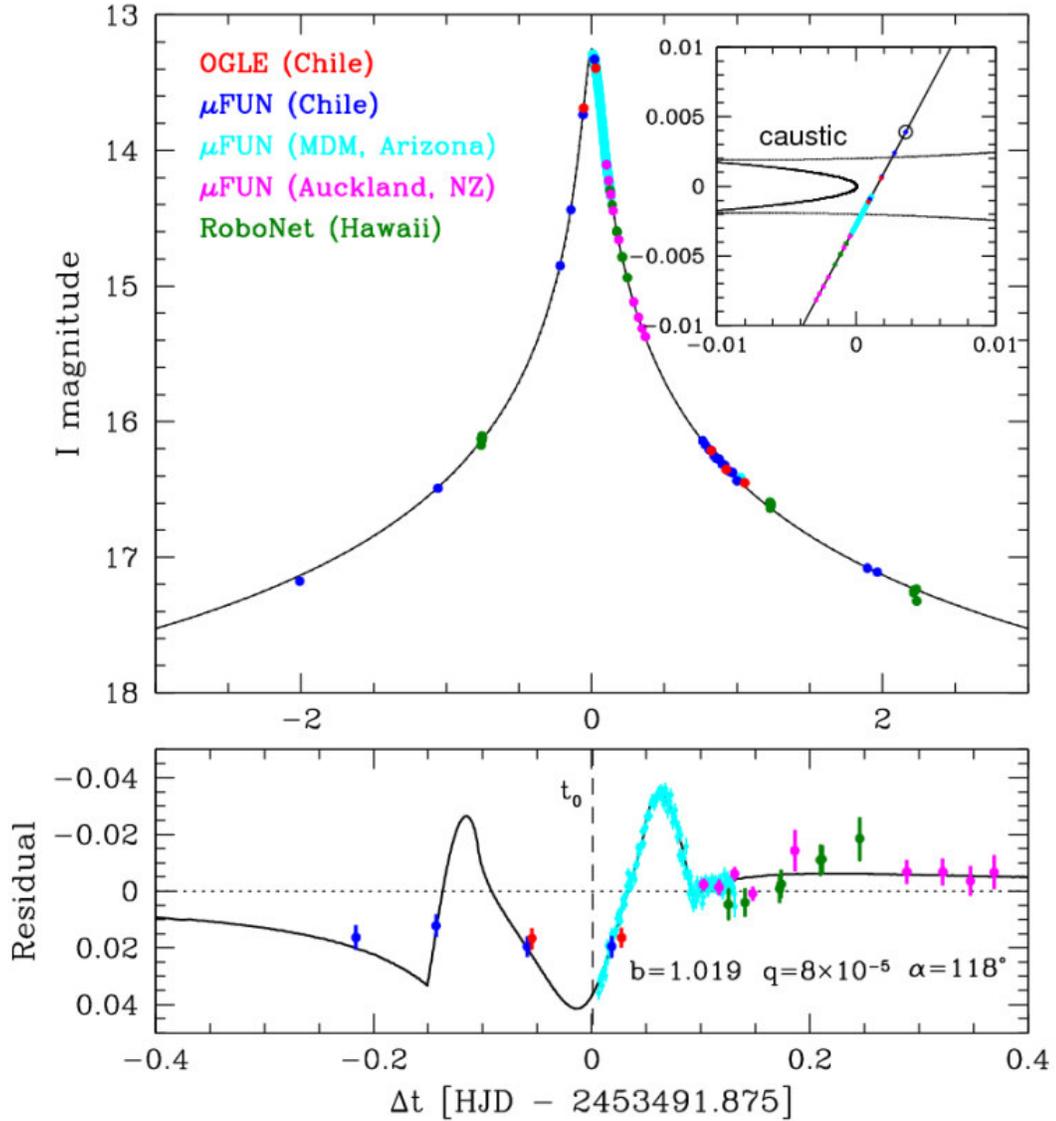

**Figure 3.** The light curve of OGLE-2005-BLG-169 is shown in the upper plot. The geometry the source star passing through the binary gravitational lens with two caustic crossings is shown in the insert. The lower plot shows the anomalous deviation of the light curve from the simple PSPL model. (Gould et al, 2006).

## 6. Observing Microlensing Events

During the 2005 microlensing season, Auckland Observatory contributed 248 hours of photometry on 27 events notified by MicroFUN (26 from OGLE, 1 from MOA). Two of these events have now been analyzed in detail, leading to the discovery two exoplanets.

The first one in April 2005 was a planet with a mass of three Jupiters in event OGLE-2005-BLG-071 (Udalski et al, 2005) - the final light curve for the event is shown in Figure 2. The peak magnification was A~70 and the planet caused anomalies which were large (0.75 magnitudes).

The second event occurred about 10 days later and detected a Neptune-mass planet (13 Earths) in event OGLE-2005-BLG-169 (Gould et al, 2006). This was a very high magnification event (A~800) and was therefore extremely sensitive to planets. Its final light curve in Figure 3 shows that the amplitude of the anomaly is about 3-4%.

A number of the other events observed in 2005 also displayed clear anomalies but these still await detailed investigation.

The first microlensing observations from Auckland were made in the 2003 at the request of the MOA collaboration. These attempts were only partly successful because it proved difficult to identify the faint target in a crowded field.

This problem was resolved by firstly improving the telescope pointing using TPoint (Software Bisque, Golden CO) and, secondly, using the "Image Link" facility available in The Sky, the program used to control the mounting.

The protocol used now is to first slew to the microlens coordinates which locates the field to ~1 arc-minute. A short exposure is taken and registered using "Image Link". Essentially, this involves extracting the position of the stars from the CCD image and matching this pattern to the stars in that vicinity listed in the stellar databases available (GSC, UCAC 2 or USNO A2.0). Image Link takes only a few seconds and exactly aligns the CCD image over the chart of the star field which includes the marked position of the microlens.

This positively identifies the microlens on the CCD frame and confirms the telescope pointing to ~1as. The telescope is re-positioned to center on the microlens and then the imaging sequence is started. Exposure times in the range 120-600s can be used without autoguiding.

Once the first full length exposure is downloaded, the field is positively identified by inspection using the annotated image provided by the survey team (OGLE or MOA). In crowded fields this procedure can be very difficult because the survey image has been taken with a much bigger telescope under much better seeing conditions. However, the fact that we already know that the microlens is located at the center of our image is a significant advantage.

Using this procedure, it now only takes a few minutes to start an observing run on any microlens after receiving an alert.

The image field of view need not be large because the fields are always crowded and there is no shortage of reference stars. We image using a 512x512 pixel subframe located at the center of the field (and thus on the optical axis); this corresponds to 11x11 arc-minutes. After calibration the images are usually cropped to 200x200 pixels to reduce the FTP upload time.

Once an imaging sequence is started, the only interruptions, apart from the weather, are to take sets of bias frames, as discussed above, or to improve the focus.

The image scale must be chosen to ensure the star images are fully sampled. The full-width half-maximum (FWHM) of a bright stellar image should be in the range of 2-3 pixels. All reduction codes used in crowded fields require the fitting of a point spread function (PSF) to each stellar image and this cannot be done accurately if the image has been under sampled. An example of a typical microlensing field is given in Figure 4.

The primary disadvantage from *over* sampling the image (i.e. FWHM > 3 pixels) is that there will be more shot noise in the stellar image than is necessary. Nevertheless, this creates fewer problems than are encountered with under sampled images.

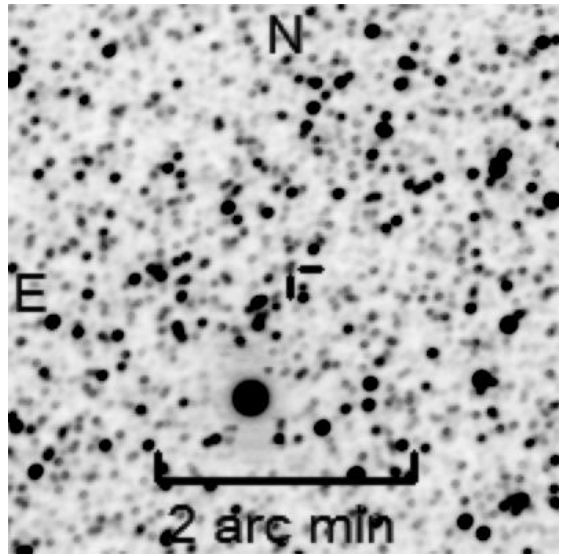

**Figure 4. The CCD image of event OGLE-2005-BLG-071 (10x120s) from Auckland Observatory.**

It is important to ensure the focus is as good as possible. Often the microlens will be blended with the images of nearby stars so the possibility of accurately disentangling the light from overlapping star images is improved if the FWHM is small (but not significantly <2 pixels).

In most microlensing events, the objective should be to maximize the total exposure time. If the microlens is bright, in which case there are likely to be caustic crossings with little advance warning, continuous imaging guarantees no critical event will be missed. If the microlens is faint then even with a small telescope, binned observations over a period of hours can still provide an accurate magnitude estimate that contributes to the coverage of the event. Successful results have even been obtained by binning 6 hours of exposure time to yield a single observation.

## 7. Image Reduction

The crowded fields of microlensing events preclude the use of aperture photometry techniques. Aperture photometry is only used at the telescope while an event is in progress to provide quick updates to other members of the follow-up network.

On completion of an observing run the Auckland images are calibrated as described above and usually cropped to 200x200 pixels. They are transmitted by FTP to OSU where they are processed through the DoPhot reduction code (Schechter, 1999). This provisional PSF photometry is then posted on the MicroFUN website and used to monitor the progress of the event.

Because of the wide differences between the various MicroFUN telescopes, detectors and filters used (or not used), this initial data set is not internally consistent. Some observatories process their own images using other reduction codes and only submit the reduced magnitudes.

If the event proceeds to the modeling phase, it is preferable, but not essential, to process all images of the event through the most accurate available image reduction pipeline. This also minimizes the possibility of any systematic differences that may due to the different reduction algorithms.

The best image reduction codes currently available use differential image subtraction (DIA) (see Alard & Lupton, 1998; Alard, 1999; Wozniak, 2000; Bond, 2001). This is the method used by the microlensing survey teams and it has been found to be close to optimal in terms of information extraction. For recent microlensing events, processing all photometry through the OGLE DIA pipeline has significantly improved the accuracy; this is especially true for images obtained with small telescopes when the microlens is faint.

## 8. Conclusion

Gravitational microlensing is developing as a powerful technique for detecting extra-solar planets. It has already achieved some notable successes with three planet discoveries so far coming from observations in the 2005 season and more are likely.

It has been found that small telescopes providing intensive coverage of high magnification events is a surprisingly effective strategy, even when operated from an urban location. The key is their ability to re-allocate all telescope time at short notice to follow an interesting event while it is near maximum brightness.

The observation of gravitational microlensing events is both technically challenging and extremely rewarding. No two events are the same so the observer has no idea what to expect. Caustic crossing events, which can only be described as photometrically spectacular, can occur at any time and often with little or no prior warning.

Importantly, planet discoveries excite the public and attract media attention to astronomy – both important considerations for a public observatory like Auckland

## 9. Acknowledgements

A travel grant provided by the MicroFUN collaboration is gratefully acknowledged, as is the support of the Auckland Observatory and Planetarium Trust Board.

I am indebted to Phil Yock (MOA) for first suggesting I should observe gravitational microlensing events.

The support and encouragement of Andrew Gould and Richard Pogge (MicroFUN) is greatly appreciated and I thank Andrzej Udalski and the OGLE team for their steady stream of interesting events.

Finally, I thank Jennie McCormick for sharing the excitement throughout.